\documentclass[aps,twocolumn,showpacs,preprintnumbers,superscriptaddress,nofootinbib]{revtex4}

\usepackage{subfigure,graphicx,epsfig,amsmath,amsfonts,amssymb,xcolor,slashed,ulem}

\newcommand{\be}{\begin{equation}}
\newcommand{\ee}{\end{equation}}
\newcommand{\ba}{\begin{eqnarray}}
\newcommand{\ea}{\end{eqnarray}}
\newcommand{\nn}{\nonumber}
\newcommand{\mev}{\textrm{ MeV}}

\begin{document}

%%%%%%%%%%%%%%%%%%%%%%%%%%%%%%%%%%%%%%%
\title{Predictions for the {\boldmath$\Lambda_b \to J/\psi ~ \Lambda(1405)$} 
decay}
\date{\today}

\author{L. Roca}
\email{luisroca@um.es}
\affiliation{Departamento de F\'isica, Universidad de Murcia, E-30100 Murcia}

\author{M. Mai}
\email{mai@hiskp.uni-bonn.de}
\affiliation{Universit\"at Bonn, Helmholtz-Institut f\"ur Strahlen- und Kernphysik (Theorie) and Bethe Center for Theoretical Physics, D-53115 Bonn, Germany}

\author{E.~Oset}
\email{oset@ific.uv.es}
\affiliation{Departamento de
F\'{\i}sica Te\'orica and IFIC, Centro Mixto Universidad de Valencia-CSIC Institutos de Investigaci\'on de Paterna, Aptdo.22085, 46071 Valencia, Spain}

\author{Ulf-G. Mei{\ss}ner}
\email{meissner@hiskp.uni-bonn.de}
\affiliation{Universit\"at Bonn, Helmholtz-Institut f\"ur Strahlen- und Kernphysik (Theorie) and Bethe Center for Theoretical Physics, D-53115 Bonn, Germany}
\affiliation{Forschungszentrum J\"ulich, Institut f\"ur Kernphysik, Institute for Advanced Simulation, and J\"ulich Center for Hadron Physics, D-52425 J\"ulich, Germany}
%%%%%%%%%%%%%%%%%%%%%%%%%%%%%%%%%%%%%%%

\begin{abstract} 
We calculate the shape of the $\pi\Sigma$ and $\bar K N$ invariant mass 
distributions in the $\Lambda_b \to J/\psi\, \pi\Sigma$ and $\Lambda_b 
\to J/\psi \,\bar K N$ decays that are dominated by the $\Lambda(1405)$ 
resonance. The weak interaction part is the same for both processes and 
the hadronization into the different meson-baryon channels in the 
final state is related by SU(3) symmetry. The most important feature 
is the implementation of the meson-baryon final-state interaction using 
two  chiral unitary models from different theoretical groups.
Both approaches give a good description of antikaon-nucleon scattering data,
the complex energy shift in kaonic hydrogen and the line shapes of $\pi 
\Sigma K$ in photoproduction, based on the two-pole scenario for the
$\Lambda (1405)$.   We find that this reaction reflects more the higher 
mass pole and we make predictions of the line shapes and relative strength 
of the meson-baryon distributions in the final state.
\end{abstract}

\maketitle

%%%%%%%%%%%%%%%%%%%%%%%%%%%%%%%%%%%%%%%%%%%%%%%%%%%%%%%%%%%%%%%%%%
\section{Introduction}
%%%%%%%%%%%%%%%%%%%%%%%%%%%%%%%%%%%%%%%%%%%%%%%%%%%%%%%%%%%%%%%%%%

The nonleptonic weak decays of charmed and bottom hadrons are turning 
into a useful tool to learn about the nature of hadrons. Although weak 
interactions violate parity and isospin, the dominance of certain 
mechanisms at the quark level induced by the topology of the mechanisms 
and the strength of the different Cabibbo-Kobayashi-Maskawa matrix 
elements, allows one to select certain decays modes that turn out to 
be sensitive to the production of some particular hadrons, see e.g. 
Refs.~\cite{Chau:1982da,Chau:1987tk,Cheng:2010vk}. In this way, 
surprises are found like the strong signal of the $f_0(980)$ in  $B^0_s$ 
decay into $J/\psi$ and $\pi^+ \pi^-$ \cite{Aaij:2011fx,Li:2011pg}, while 
no signal was found for the $f_0(500)$. This is surprising since the 
$f_0(500)$ couples more strongly to $\pi^+ \pi^-$ than the $f_0(980)$. 
Further, in the decay of $\bar B^0$ into $J/\psi$ and $\pi^+ \pi^-$ 
\cite{Aaij:2013zpt}, the $f_0(500)$ signal was prominent  while the 
$f_0(980)$ production was strongly suppressed. Attempts to 
explain these features in terms of tetraquark structures for the 
scalar mesons were made in \cite{stone}. A different line of investigation 
has been opened in \cite{weihong} following the findings of chiral 
unitary theory, where these scalar mesons are dynamically generated 
from the interaction of pseudoscalar mesons  
\cite{npa,ramonet,kaiser,markushin,juanito,rios}. In this approach, 
the basic mechanism at the quark level is identified as follows: 
one $c \bar c$ state forms the $J/\psi$, another $q \bar q$ pair 
hadronizes into a pair of mesons, and the final state interaction 
of these mesons leads to the production of the scalar resonances. 
It should be mentioned that the use of unitarized chiral perturbation
theory to explore the physics of heavy meson decays was pioneered
in Refs.~\cite{Meissner:2000bc,Gardner:2001gc}, and has been
recently employed to quantify the S-wave pollution in semi-leptonic
B decays \cite{Doring:2013wka} and to facilitate the extraction
of $|V_{ub}|$ from $B_{\ell 4}$ decays \cite{Meissner:2013pba}.

The method of Ref.~\cite{weihong} has allowed one to interpret many other 
different decays. In this sense, ratios for the production of $J/\psi$ and 
vector mesons in $B$ decays were evaluated in \cite{Bayar:2014qha} and 
predictions for the $J/\psi \kappa(800)$ decay were also made. 
In \cite{daid} the  $D^0$ decays into $K^0_s$ and $f_0(500)$, $f_0(980)$, $a_0(980)$ 
were described.  Dynamically generated states from the vector-vector interaction 
were investigated in the $\bar{B}^0$ and $\bar{B}^0_s$ decays into $J/\psi$ 
plus $f_0(1370),~f_0(1710),~f_2(1270),~f'_2(1525),~K^*_2(1430)$ \cite{xievec}. 
Similarly,  the $\bar B^0$ decay into $D^0$ and $\rho$ or $f_0(500)$, $f_0(980)$, 
$a_0(980)$ and $\bar B^0_s$ decays into $D^0$ and $K^{*0}$ or $\kappa(800)$ were 
addressed in \cite{xiebd}. Further work is done in \cite{marina}, where the $KD$ 
scattering and the $D_{s0}^*(2317)$ resonance were studied from the  $B^0_s$ 
decay into $D_s~ DK$.  Also, semileptonic $B_s$ and $B$ decays are addressed 
in \cite{fernando}. 

In the present work we would like to follow this same line of reasoning 
but involving baryons rather than mesons.  The reaction we study here is 
$\Lambda_b \to J/\psi ~ \Lambda(1405)$, where the $\Lambda(1405)$ is to be 
seen in the $\pi \Sigma$ spectrum. This reaction is not measured yet but 
the related process $\Lambda_b \to J/\psi ~ \Lambda(1115)$ has already been 
measured by the D0   \cite{Abazov:2007sf} and the ATLAS \cite{Aad:2014iba} 
collaborations. Further, there is experimental information on the $\Lambda_b 
\to J/\psi ~ K^- p$ decay channel from the LHCb \cite{Aaij:2014zoa} and 
CDF \cite{Aaltonen:2014vra} collaborations. No absolute values are provided for 
this latter decay and only ratios to other reactions are studied. Our work will 
allow us to relate the $\Lambda_b \to J/\psi ~ \Lambda(1405)$ decay to the 
$\Lambda_b \to J/\psi ~ K^- p$ decay and ratios between the invariant 
$K^- p$ and $\pi \Sigma$ mass distributions will be provided. 

The reason to suggest the measurement of the $\Lambda(1405)$ in 
the  $\Lambda_b$ decay 
is the relevance of the $\Lambda(1405)$ as the most significant example of a 
dynamically generated resonance. Indeed, very early it was already suggested 
that 
this resonance should be a molecular state of $\bar K N$ and $\pi \Sigma$ 
\cite{Dalitz:1960du,Dalitz:1967fp}. This view has been also invoked in 
Ref.~\cite{Veit:1984an}. 
However, it was with the advent of chiral unitary theory, this idea gained 
strength \cite{Kaiser:1995eg,Kaiser:1996js,Oset:1998it,Oller:2000fj,Lutz:2001yb,
Oset:2001cn,Hyodo:2002pk,cola,GarciaRecio:2002td,GarciaRecio:2005hy,Borasoy:2005ie,
Oller:2006jw,Borasoy:2006sr,hyodonew,Mai:2012dt}.

One of the surprises of these works is that two poles were found for 
the $\Lambda(1405)$ \footnote{In fact, one might thus speak of two $\Lambda(1405)$ particles.}. 
The existence of two states was hinted in \cite{Fink:1989uk}, 
using the chiral quark model, and it was found in \cite{Oller:2000fj} using 
the chiral unitary approach. A thorough search was conducted in \cite{cola} by 
looking at the breaking of SU(3) in a gradual way, confirming the existence of 
these two poles and its dynamical origin. One of the consequences of this two-pole 
structure is that the peak of the resonance does not always appear at the same energy, 
but varies between 1420~MeV and 1480~MeV depending on the reaction used 
\cite{Thomas:1973uh,Hemingway:1984pz,Niiyama:2008rt,prakhov,Moriya:2012zz,Moriya:2013eb, Zychor:2007gf,fabbietti}. 
This is because different reactions give different weights to each of the 
poles. While originally most reactions gave energies around 1400~MeV, 
the origin of the nominal mass of the resonance,  the $K^- p \to \pi^0\pi^0 \Sigma^0$
was measured \cite{prakhov} and a peak was observed around 1420~MeV, 
narrower than the one observed in \cite{Thomas:1973uh,Hemingway:1984pz}, which 
was interpreted within the chiral unitary approach in \cite{magaslam}. Another 
illustrating experiment was the one of \cite{Braun:1977wd}  where  a clear 
peak was observed around 1420~MeV in the $K^- d \to n \pi \Sigma$ reaction, 
which was also interpreted theoretically in \cite{sekihara} along the same lines, see also Refs.~\cite{Miyagawa:2012xz,Jido:2012cy}. Very recently it has 
also been suggested that the  neutrino induced production 
of the $\Lambda(1405)$ is a good tool to further investigate the properties and 
nature of this resonance \cite{Ren:2015bsa}. 

The basic feature in the dynamical generation of the $\Lambda(1405)$ in 
the chiral unitary approach is the coupled channel unitary treatment of the 
interaction between the coupled channels $K^-p$, $\bar{K}^0n$, $\pi^0\Lambda$, 
$\pi^0\Sigma^0$, $\eta\Lambda$, $\eta\Sigma^0$, $\pi^+\Sigma^-$, $\pi^-\Sigma^+$, $K^+\Xi^-$ 
and $K^0\Xi^0$. The coupled channels study allows us to relate the $K^-p$ and $\pi \Sigma$  
production, where the resonance is seen, and this is a unique feature of the nature of 
this resonance as a dynamically generated state. It allows us to make
predictions for the $\Lambda(1405)$ production from the measured 
$\Lambda_b \to J/\psi ~ K^-p$ decay.

Technically, the work proceeds as follows: the basic mechanism for the 
$\Lambda_b \to J/\psi ~ K^- p$ decay at the quark level is identified. First,
a $c \bar c$ state is produced, which forms the the 
$J/\psi$,  and the three remaining light quarks $u,d,s$ hadronize 
to a meson-baryon pair. After this, the latter undergoes final 
state interactions in coupled channels, such that  the $\Lambda(1405)$ is 
unavoidably produced. To calculate the corresponding decays, 
we shall use two different  models  of the coupled channels interaction: 
One of them \cite{Roca:2013av,Roca:2013cca} 
uses the lowest order chiral Lagrangians slightly modified to fit the 
photoproduction data from CLAS \cite{Moriya:2012zz,Moriya:2013eb}. 
The other one 
incorporates explicitly the next-to-leading order Lagrangian with coefficients 
that are also fitted to the same data \cite{Mai:2014xna}. This latter approach
has been used to generate theoretical uncertainties, which are important for
judging the precision achieved. In spite of the apparent 
differences, the results for different observables are remarkably similar in 
both approaches and the two poles obtained are practically identical and 
quite similar to those obtained in \cite{cola}.

This is the first theoretical work done for this reaction, yet it shares 
some aspects with a similar process, the $\Lambda_c \to \pi^+ \pi^- \Sigma$ 
reaction,  which was proposed in \cite{Hyodo:2011js} as a tool to measure 
the  $\pi^- \Sigma$  scattering  length. Indeed, in \cite{Hyodo:2011js} 
the hadronization of the final 
three quark state at the tree level is done, albeit in a different way, and the 
final-state interaction of coupled channels is described in a similar
manner as done here. Other 
works for related reactions use  quark models to evaluate amplitudes, 
like in the study of the  $\Lambda_b \to J/\psi ~ \Lambda(1115)$ reaction 
\cite{Gutsche:2013oea}, or the semileptonic transitions from 
$\bar B_s \to K l \bar \nu_l$ \cite{Albertus:2014gba}. Further, some studies
make use of  heavy quark effective theory to evaluate related amplitudes as 
for  the process $\Lambda_b \to \Lambda_c l \bar \nu_l$ \cite{Jia:2012vt}. In 
contrast to these later works, the one presented here, as well as the 
one of \cite{Hyodo:2011js}, does not perform a microscopic study of the 
reaction since we do not aim at obtaining absolute rates, instead we exploit 
the dynamics of the coupled channels to relate the distributions of invariant 
masses in different final states, hopefully contributing
to a better understanding of the meson-baryon interaction and the nature 
of some resonances, in particular the $\Lambda(1405)$.

%%%%%%%%%%%%%%%%%%%%%%%%%%%%%%%%%%%%%%%%%%%%%%%%%%%%%%%%%%%%%%%%%%
\section{Formalism}\label{sec:formalism}
%%%%%%%%%%%%%%%%%%%%%%%%%%%%%%%%%%%%%%%%%%%%%%%%%%%%%%%%%%%%%%%%%%

\begin{figure*}[t]
%%%%%%%%%%%%%%%%%
\begin{minipage}{0.48\linewidth}
 \includegraphics[width=\linewidth]{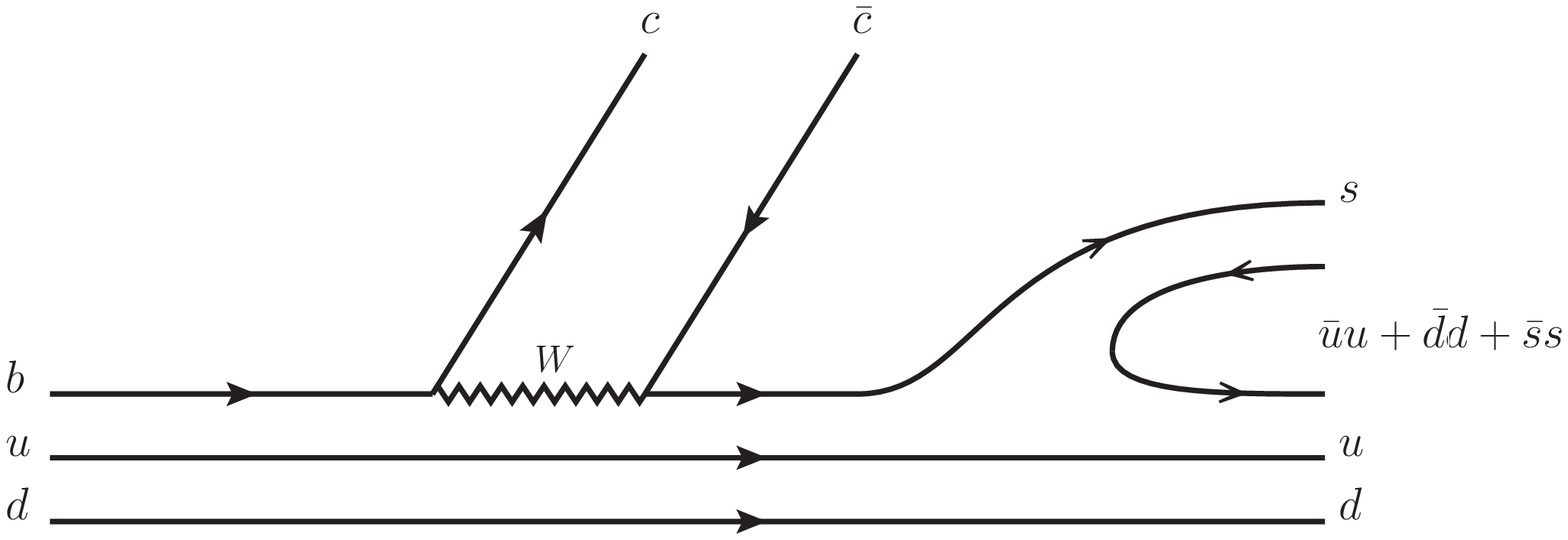}
$\hspace{0.05\linewidth}
\underbrace{\hspace{0.5\linewidth}}_{\text{Weak decay}}
\underbrace{\hspace{0.35\linewidth}}_{\text{Hadronization}}
\hspace{0.1\linewidth}$
\caption{Production of a $K^-p$ pair from the weak decay 
${\Lambda_b\to\Lambda\,J/\psi}$ 
via a hadronization mechanism. The full and wiggly lines correspond to 
quarks and the $W$-boson, respectively.}\label{fig:weak}
\end{minipage}
%%%%%%%%%%%%%%%%
~~~
\begin{minipage}{0.48\linewidth}
\vspace{+1cm}
\includegraphics[width=\linewidth]{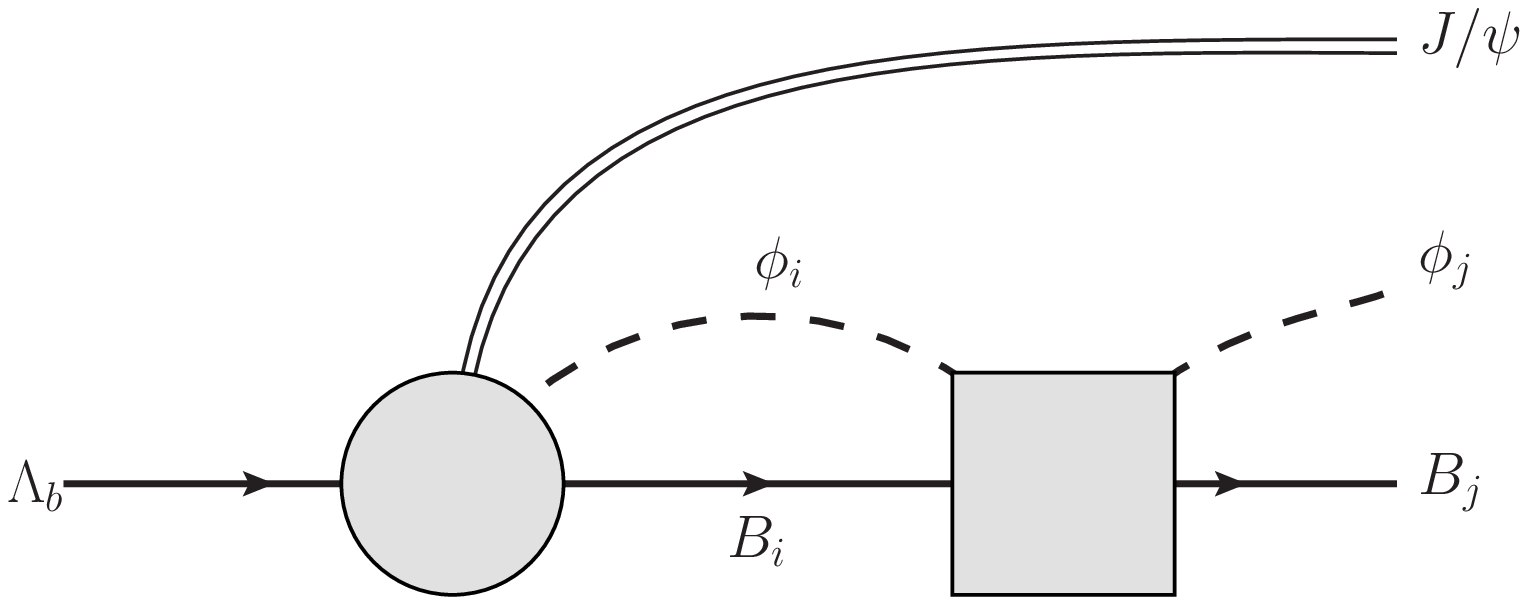}
\vspace{-0.1cm}
\caption{Final-state interaction of the meson-baryon pair, where the double, 
full and dashed lines denote 
the $J/\psi$, the baryons and the pseudoscalar mesons, respectively. 
The circle and square denote the production mechanism of the $J/\psi B_i\phi_i$ 
as depicted in Fig.~\ref{fig:weak} as well as meson-baryon scattering 
matrix $t_{ij}$, respectively}\label{fig:full}
\end{minipage}
%%%%%%%%%%%%%%%%
\end{figure*}

In this section we describe the reaction mechanism for the process 
$\Lambda_b \to J/\psi\,\Lambda(1405)$, which is divided into three parts. 
The first  two parts describe the decay mechanism $\Lambda_b\to J/\psi\,B\phi$, 
with $B\phi$ the meson-baryon system of strangeness $S=-1$, in the language of 
the quark model. Then, after hadronization, 
the final-state interaction is described in terms of the 
effective (hadronic) degrees of freedom of chiral perturbation theory (ChPT). 
After a resummation of the chiral meson-baryon potential to an infinite order, 
the $\Lambda(1405)$ is generated dynamically. In the following, we describe each 
single step of this reaction mechanism in more detail.

\smallskip

\noindent\textbf{Weak decay:} The $b$ quark of the $\Lambda_b$ undergoes the 
weak transition to a $c\bar c$ pair and an $s$-quark as depicted in the left 
part of Fig.~\ref{fig:weak}. This transition is quantified by the matrix elements 
of the Cabibbo-Kobayashi-Maskawa (CKM) matrix $V_{cb}V_{cs}^*$ and it 
is favored compared to $b\to c\bar c d$ leading to the
$\Lambda_b\to J/\psi p\pi^-$, which was observed for the first time by the 
LHCb collaboration, see Ref.~\cite{Aaij:2014zoa}.

\smallskip

\noindent\textbf{Hadronization:} The $c \bar c$ pair forms the well-known 
$J/\psi$, while the virtual $uds$ three quark state  undergoes  hadronization 
to form a meson-baryon pair. This happens due to the large phase space 
available ($\leq 2522$~MeV for $M_{\Lambda_b}=5619$~MeV, 
$M_{J/\psi}=3097$~MeV), so that a quark-antiquark pair can become real, forming 
together with the three available quarks a meson-baryon pair.
In principle, different meson-baryon states can be produced in such a mechanism. 
To determine their  relative significance, we assume first that the $u$ and $d$ 
quarks of the original $\Lambda_b$ state are moving independently in a potential well. 
Further, we note that the $\Lambda_b$ ($J^p=1/2^+$)  is in the ground state of the 
three-quarks $(udb)$. Therefore, all relative angular momenta 
between different quarks are zero. After the weak transition, but before the 
hadronization, the three-quark state $(uds)$ has to be in a p-wave since the 
final $\Lambda(1405)$ is a negative-parity 
state. On the other hand, since the $u$ and $d$ quarks are considered to  
be spectators and they were originally in $L=0$, the only possibility is that 
the $s$ quark carries the angular momentum, $L=1$. Moreover, since the final 
mesons and baryons are in the ground state and in s-wave to each other,
all the angular momenta in the final state are zero. Consequently, the $\bar q q$ 
pair cannot be produced elsewhere, but between the $s$ quark and the $ud$ pair 
as depicted in Fig.~\ref{fig:weak}.

The flavour state of the initial $\Lambda_b$ can be written as
\begin{align*}
|\Lambda_b\rangle=\frac{1}{\sqrt{2}}|b(ud-du)\rangle\,,
\end{align*}
turning after the weak process into
\begin{align*}
\frac{1}{\sqrt{2}}|s(ud-du)\rangle\,,
\end{align*}
since the $u$ and $d$ quarks are considered to be spectators. Thus, after 
hadronization, the  final quark flavor state is
\begin{align*}
|H\rangle&\equiv \frac{1}{\sqrt{2}}|s\,(\bar u u +\bar d d +\bar s s)\,(ud-du)\rangle\\
         &=\frac{1}{\sqrt{2}} \sum_{i=1}^3{|P_{3i}q_i(ud-du)}\rangle\,,
\end{align*}
where we have defined
\begin{equation*}
q\equiv \left(\begin{array}{c}u\\d\\s\end{array}\right)\,\text{~~and~~~}
P\equiv q\bar q^\tau=\left(\begin{array}{ccc}u\bar u & u\bar d & u\bar s\\
					       d\bar u & d\bar d & d\bar s\\
					       s\bar u & s\bar d & s\bar s
			      \end{array}\right)\,.
\end{equation*} 
The latter is nothing else than the quark-antiquark representation of the $SU(3)$
pseudoscalar meson matrix
\begin{eqnarray*}
P=
\left(\begin{array}{ccc} 
              \frac{\pi^0}{\sqrt{2}}  + \frac{\eta}{\sqrt{3}}+\frac{\eta'}{\sqrt{6}}& \pi^+ & K^+\\
              \pi^-& -\frac{1}{\sqrt{2}}\pi^0 + \frac{\eta}{\sqrt{3}}+ \frac{\eta'}{\sqrt{6}}& K^0\\
              K^-& \bar{K}^0 & -\frac{\eta}{\sqrt{3}}+ \frac{2\eta'}{\sqrt{6}} 
      \end{array}
\right)\,,
\end{eqnarray*}
where we have assumed the ordinary mixing
between the singlet and octet $SU(3)$ states
for the the $\eta$ and $\eta'$, see e.g. Ref.~\cite{Bramon:1992kr}:
\begin{eqnarray*}
\eta = \frac{1}{3}\eta_1 + \frac{2\sqrt{2}}{3}\eta_8\,, \qquad
\eta' = \frac{2\sqrt{2}}{3}\eta_1 - \frac{1}{3}\eta_8\,.
\end{eqnarray*}
The hadronized state $|H\rangle$ can now be written as
\begin{align*}
|H\rangle= \frac{1}{\sqrt{2}} \bigg( K^- u(ud&-du)+\bar K^0 d(ud-du)\nn \\
         &+\frac{1}{\sqrt{3}}\left(-\eta+\sqrt{2}\eta'\right)s(ud-du)\bigg).
\end{align*}
We can see that these states have overlap with the mixed antisymmetric baryon state
\cite{close_book}. Further, the flavour states of the final 
octet baryons can be written as
\begin{align*}
|p      \rangle=&\frac{1}{\sqrt{2}}|u(ud-du)\rangle\,,\\
|n      \rangle=&\frac{1}{\sqrt{2}}|d(ud-du)\rangle\,,\\
|\Lambda\rangle=&\frac{1}{\sqrt{12}}| (usd-dsu)+(dus-uds) +2(sud-sdu)\rangle\,.
\end{align*}
Consequently, the hadronized state can be expressed in terms of ground state
octet mesons and baryons as 
\begin{equation}
|H\rangle=|K^-p\rangle+|\bar K^0 n\rangle
-\frac{\sqrt{2}}{3}|\eta\Lambda\rangle+\frac{2}{3}|\eta'\Lambda\rangle\,,
\label{eq:Hflav}
\end{equation}
which provides the relative weights between the final meson-baryon channels.
Note that there is not direct production of $\pi\Sigma$ and $K\Xi$, however,
these channels are present through the intermediate loops in the final state 
interaction as described below. Moreover, the final $\eta'\Lambda$ channel 
will be neglected since it has a small effect due its high mass and can be 
effectively reabsorbed in the regularization parameters as will be 
explained below.

\smallskip

\noindent\textbf{Formation of the {\boldmath$\Lambda(1405)$}:} After the production of  
a meson-baryon pair, the final-state interaction takes place, which is 
parametrized by the scattering matrix
$t_{ij}$. Thus, after absorbing the CKM matrix elements and kinematic prefactors into an 
overall factor $V_p$, the amplitude $\mathcal{M}_{j}$ for the transition 
$\Lambda_b\to J/\psi\,\phi_jB_j$ can be written as
\begin{align}\label{eqn:fullamplitude}
\mathcal{M}_{j}(M_{\rm inv})=V_p\left( h_j+\sum_{i}h_iG_i(M_{\rm inv})\,t_{ij}(M_{\rm inv}) \right)\,,
\end{align}
where, considering Eq.~(\ref{eq:Hflav}),
\begin{align*}
&h_{\pi^0\Sigma^0}=h_{\pi^+\Sigma^-}=h_{\pi^-\Sigma^+}=0\,,~h_{\eta\Lambda}=
-\frac{\sqrt{2}}{3}\,,\\
&h_{K^-p}=h_{\bar K^0n}=1\,,~h_{K^+\Xi^-}=h_{K^0\Xi^0}=0\,,
\end{align*}
and $G_i$ denotes the one-meson-one-baryon loop function, chosen in 
accordance with the models for the scattering matrix\footnote{More precisely,
$t_{ij}$ denotes the s-wave contribution to the scattering matrix.} $t_{ij}$ 
as it will be described below. Further, $M_{\rm inv}$
is the invariant mass of the meson-baryon system in the final state. 
Note also that the above
amplitude holds for an s-wave only and every intermediate particle is put on the
corresponding mass shell. Finally, the invariant mass distribution $\Lambda_b\to
J/\psi\,\phi_jB_j$ reads
\begin{align}\label{eqn:dGammadM}
\frac{d\Gamma_j}{dM_{\rm inv}}(M_{\rm inv})
=\frac{1}{(2\pi)^3}\frac{m_j}{M_{\Lambda_b}}{\rm\textbf{p}}_{J/\psi} {\rm\textbf{p}}_j\left|\mathcal{M}_{j}(M_{\rm inv})\right|^2\,,
\end{align}
where $\rm\textbf{p}_{J/\psi}$ and $\rm\textbf{p}_j$ denote the modulus of the 
three-momentum
of the $J/\psi$ in the $\Lambda_b$ rest-frame and the modulus of the center-of-mass
three-momentum in the final meson-baryon system, respectively. The mass of the
final baryon is denoted by $m_j$.

As already described in the introduction, the baryonic $J^P=1/2^-$ resonance 
$\Lambda(1405)$
has to be understood as a dynamically generated state from the coupled-channel effects. The
modern approach for it is referred to as chiral unitary models, see e.g.
Refs.~\cite{Kaiser:1995eg, Kaiser:1996js, Oset:1998it, Oller:2000fj,Lutz:2001yb, Oset:2001cn,
Hyodo:2002pk,cola, GarciaRecio:2002td,GarciaRecio:2005hy, Borasoy:2005ie, Oller:2006jw,
Borasoy:2006sr, hyodonew, Mai:2012dt}. In the present approach we use the scattering
amplitude from two very recent versions of such approaches, see
Refs.~\cite{Roca:2013av,Roca:2013cca,Mai:2014xna}. While the basic motivation 
is the same for both approaches there are several important differences, which 
shall be described in the following two subsections.

%%%%%%%%%%%%%%%%%%%%%%%%%%%%%%%%%%%%%%%%%%%%%%%%%%%%%%%%%%%%%%%%%%
\subsection{Summary of the Bonn model}\label{sec:mmmodel}
%%%%%%%%%%%%%%%%%%%%%%%%%%%%%%%%%%%%%%%%%%%%%%%%%%%%%%%%%%%%%%%%%%

The model described in the present subsection has been developed originally 
in Ref.~\cite{Bruns:2010sv} and used  first for the analysis of the 
lowest $S_{11}$ nucleon resonances from scattering data as well as single-meson 
photoproduction data in Ref.~\cite{Mai:2012wy}. Later in  Ref.~\cite{Mai:2012dt} 
it was also applied to meson-baryon scattering in the strangeness $S=-1$ sector, 
adjusting the free parameters of the model to the available scattering data (including
the threshold data from kaonic hydrogen).
While the data was described quite satisfactorily, the broad pole of the 
$\Lambda(1405)$ appeared at a different position than usually assumed. While 
the reason for this discrepancy may have various roots, see the discussion in 
Ref.~\cite{Mai:2012dt}, one important systematic observation was made there. 
Namely,  the off-shell contributions of the intermediate particle fields in the 
Feynman diagrams are quite moderate in this setting. This observation is enormously 
useful as it allows to reduce the computational effort by a factor of 30-60 and  
therefore to study the large parameters space of this model in more detail as it 
was done in Ref.~\cite{Mai:2014xna}. There, in a large scale analysis of the 
parameter space we have found several solutions including similar ones to that 
of Ref.~\cite{Mai:2012dt}. However, in a very conservative test against the recent 
and very precise two-meson photoproduction data by the CLAS collaboration 
\cite{Moriya:2012zz,Moriya:2013eb} many solutions were ruled out. The best 
solution of this procedure is used here. In what follows we will describe the 
major features of this approach, while for details the reader is referred 
to  Refs.~\cite{Mai:2012dt,Bruns:2010sv,Mai:2014xna}.

The driving term of this model is the chiral potential, derived from the 
leading and next-to-leading order 
chiral Lagrangian in the three flavour formulation, see Ref.~\cite{Frink:2004ic}. 
In the on-shell approximation,  this potential reads
\begin{equation}
V(\slashed{p})=A(p^2)+B(p^2)\slashed{p}~,
\end{equation}
with
\begin{align*}
A(p^2)=\Big(&-A_{WT}(m_{i}+m_{f})+A_{14}(q_{i}\cdot q_{f})\nonumber\\
      &+2A_{57}\big((q_{i}\cdot q_{f})-p^2-m_{i}m_{f}\big)\nonumber\\
      &-A_{811}\big(m_{f}(q_{i}\cdot p)+m_{i}(p\cdot q_{f})\big)+A_M      \Big)\,,\nonumber\\
B(p^2)=\Big(&2A_{WT}\nonumber\\
      &+2A_{57}(m_{i}+m_{f})+A_{811}\big((q_{i}\cdot p)+(p\cdot q_{f})\big)\Big)\,,\nonumber    
\end{align*}
where here and in the following $M/m$ and $q/p$ denote the meson/baryon mass and 
the meson/overall four-momentum, 
respectively, with $p^2 = M_{\rm inv}^2$. The index $i/f$ denotes the in-/out-going 
states.  The $A_{WT}$, $A_{14}$, $A_{57}$, $A_{M}$ and $A_{811}$ are 10-dimensional 
matrices which encode the coupling strengths between all 10 channels of the meson-baryon system for strangeness $S=-1$, i.e. $\{K^-p$, 
$\bar K^0 n$, $\pi^0\Lambda$, $\pi^0\Sigma^0$, $\pi^+\Sigma^-$, $\pi^-\Sigma^+$, $\eta\Lambda$, $\eta \Sigma^0$, 
$K^+\Xi^-$, $K^0\Xi^0\}$. They are given explicitly 
in Ref.~\cite{Mai:2014xna}. Setting all meson masses and decay constants to their 
physical values, the only unknown of the above equation are the 14 low-energy 
constants (LECs) of SU(3) ChPT at NLO. These LECs serve as free parameters of the 
present model as they are not known precisely at the moment. 

At any finite order, the strict chiral expansion of the scattering amplitude in 
the baryon sector is restricted to a certain range around the point $p^2=m_0^2$ 
(with $m_0$ the octet mass in the chiral limit) and a small 
momentum transfer to the baryon. Moreover, at any finite order such a series fails in 
the vicinity of resonances such as the $\Lambda(1405)$, located just below the 
$\bar K N $ threshold. Therefore, a resummation of the driving term is required 
to describe this system. In the present work we use the coupled-channel Bethe-Salpeter 
equation in the on-shell approximation. Here, the 
scattering amplitude $T(\slashed{p})$ is the solution of the following matrix equation over the $10$-dimensional channel space
\begin{align}\label{eqn:Bonn-BSE}
T(\slashed{p})=V(\slashed{p})+V(\slashed{p})\,G(M_{\rm inv})\,T(\slashed{p})\,,
\end{align}
where $G$ is a diagonal matrix, containing the one-meson-one-baryon loop functions as elements, which on-shell read
\begin{align}\label{eq:bonnG}
G^{ij}(M_{\rm inv})=i\int\frac{d^dl}{(2\pi)^d}\frac{2m_i\delta^{ij}}{(l^2-M_i^2+i\epsilon)((l-p)^2-m_i^2+i\epsilon)}\,.
\end{align}
This function is treated in dimensional regularization, applying the usual $\overline{\rm MS}$
subtraction scheme. It should be noted that due to the non-perturbative character of
Eq.~(\ref{eqn:Bonn-BSE}) the regularization scale is treated as a free parameter 
of the model. In the isospin basis, there are 6 such parameters. 
All free parameters of the model are taken from the solution \#4 from 
Ref.~\cite{Mai:2014xna}, which was found to be the best solution, describing 
all available meson-baryon scattering data as well as the recent two-meson 
photoproduction data by the CLAS collaboration \cite{Moriya:2012zz,Moriya:2013eb}. 
For the purpose of the present work, the scattering amplitude $T(\slashed p)$ of 
this solution is projected to the lowest partial wave, i.e. $f_{0+}$. The latter is 
related to the scattering matrix $t_{ij}$ from the Eq.~(\ref{eqn:fullamplitude}) via
\begin{align}\label{eq:norm}
t^{ij}(M_{\rm inv})=-\frac{4\pi M_{\rm inv}}{\sqrt{m_i m_j}}f_{0+}^{ij}(M_{\rm inv})\,.
\end{align}
For completeness, we recall that two poles of $\Lambda(1405)$ were found for 
this solution, located on the second Riemann sheet connected to the first one 
between the $\pi\Sigma$ and $\bar KN$ thresholds. Their positions are  
$({1429^{+8}_{-7}-i\,12^{+2}_{-3}})$~MeV and $({1325^{+15}_{-15}-i\,90^{+12}_{-18}})$~MeV.
Here, the error bars are due to fit parameter errors. Naturally, the latter lead to an 
uncertainty of the scattering amplitude $t^{ij}(M_{\rm inv})$ which is discussed in detail
in Ref.~\cite{Mai:2014xna}. The focus of the present work lies on the 
the systematic error, considering two different models for 
the final-state interactions in the $\Lambda_b$ decay, and 
we will omit these parameter errors in what follows.

%%%%%%%%%%%%%%%%%%%%%%%%%%%%%%%%%%%%%%%%%%%%%%%%%%%%%%%%%%%%%%%%%%
\subsection{Summary of the MV-model  }\label{sec:ormodel}
%%%%%%%%%%%%%%%%%%%%%%%%%%%%%%%%%%%%%%%%%%%%%%%%%%%%%%%%%%%%%%%%%%

Let us  briefly review the second unitarized meson-baryon model 
\cite{Roca:2013av,Roca:2013cca} that we are going to use in the present work
(which we will call MV-model, after Murcia-Valencia, in the following), 
for the sake of
completeness and to ease the comparison with the Bonn-model
summarized in section~\ref{sec:mmmodel}. The aim of the studies carried out in
Refs.~\cite{Roca:2013av,Roca:2013cca} was to fine tune the meson-baryon scattering 
amplitudes obtained in the chiral unitary approach  by allowing to change slightly 
the unitarization kernel and loop functions
through the inclusion of free parameters of natural order which were fitted 
to the $\gamma p \to K^+ \pi\Sigma$  data from CLAS. 

The basic model for the unitarized meson-baryon scattering amplitude has 
been widely developed and applied in many previous works (see for instance 
\cite{Oset:1998it,Oller:2000fj,Hyodo:2006kg,Oset:2001cn}). The
chiral unitary approach is based on the implementation of unitarity and 
the exploitation of the analytic
properties of the scattering amplitudes with the only input of the lowest orders 
chiral potentials. This
has been usually  carried out by means of the Inverse Amplitude Method \cite{dobado-pelaez,ramonet} or the
N/D method \cite{Oller:1998zr,Oller:2000fj,Hyodo:2003qa} which was shown 
in \cite{npa} to be equivalent to
the Bethe-Salpeter equation. From the $N/D$ method, the scattering amplitude 
$t_{ij}$ fulfills
Eq.~(\ref{eqn:Bonn-BSE}) which provides the solution
\begin{equation}
t=[1-vG]^{-1}v\,,
\label{eq:BS}
\end{equation}
in the normalization of Eq.~(\ref{eq:norm}), with $v_{ij}$ the $s$-wave projected 
meson-baryon potential described below, Eq.~(\ref{eq:WT}).

In the MV-model, the interaction kernel $v_{ij}$ obtained from the lowest order chiral  Lagrangian for the
interaction of the octet of pseudoscalar mesons with the octet of the lowest mass $1/2^+$ baryons
\cite{Bernard:1995dp}. The $s$-wave projected potential reads~\cite{Oset:2001cn}
\begin{align}
v_{ij}(M_{\rm inv})=-C_{ij}&\frac{1}{4f^2}(2M_{\rm inv}-m_i-m_j) \nonumber\\
\times&\left(\frac{m_i+E_i}{2m_i}\right)^{1/2}
\left(\frac{m_j+E_j}{2m_j}\right)^{1/2}\,, \label{eq:WT}
\end{align}
where $f$ is the averaged meson decay constant $f = 1.123f_\pi$~\cite{Oset:2001cn} with $f_\pi = 92.4$~MeV, $E_i$ ($m_i$) the energies (masses) of the baryons of the $i$-th channel and the $C_{ij}$ are coefficients, that  
for isospin $I=0$ are given by 
\begin{equation}
C_{ij} =\begin{pmatrix} 
        3 & -\sqrt{\frac{3}{2}}  \\ 
        -\sqrt{\frac{3}{2}}& 4 
        \end{pmatrix} \,,
\label{eq:couplingI}
\end{equation}
where the  $i$ and $j$ subscripts stand for $\bar K N$ and $\pi\Sigma$ in 
isospin-basis. Eqs.~(\ref{eq:WT}) and (\ref{eq:couplingI})
 represent the standard Weinberg-Tomozawa interaction, slightly modified to incorporate 
 relativistic corrections \cite{Oset:2001cn}. Note that we work in an isospin symmetric formalism for the 
meson-baryon interaction. Further, the values of the elements of the matrix
$C_{ij}$ are given by chiral symmetry.
The other meson-baryon
channels in $I=0$ and strangeness $S=-1$,  $\eta\Lambda$ and $K\Xi$, are not explicitly included. Indeed, since  the thresholds of these channels lay far above from the  energies that we will consider in the
present work, they can effectively be reabsorbed in the regularization parameters that we will explain
below.
%%%%%%%%%%%%%%%%%%%%%%%%%%%%%%%%%%%%%%%%%%%%%%%%%%%%%%%%%%%%%%%%%%%
% \begin{figure}[thb]
% \begin{center}
% \includegraphics[width=\linewidth]{figure3.eps}
% \caption{Modulus squared of the $I=0$ meson-baryon unitarized
% amplitudes $t^{I=0}_{\bar K N,\bar K N}$ (solid line), $t^{I=0}_{\pi\Sigma,\pi\Sigma}$ (dashed line), 
% $tT^{I=0}_{\bar K N,\pi\Sigma}$ (dashed-dotted line).
% Note the different shape of the amplitudes as a consequence of the 
% different weight of the two poles.
% }
% \label{fig:t_MMMM0fit}
% \end{center}
% \end{figure}
%%%%%%%%%%%%%%%%%%%%%%%%%%%%%%%%%%%%%%%%%%%%%%%%%%%%%%%%%%%%%%%%%%%%%

In refs.~\cite{Roca:2013av,Roca:2013cca}  the coefficient matrix $C_{ij}$, 
Eq.~(\ref{eq:couplingI}), was substituted by 
\begin{equation} 
C_{ij} =  \begin{pmatrix} 
          3 \alpha_{11} & -\sqrt{\frac{3}{2}} \alpha_{12}  \\ 
          -\sqrt{\frac{3}{2}}\alpha_{12}& 4 \alpha_{22}
\end{pmatrix}\,,\label{eq:couplingIalpha}
\end{equation}
where the parameters $\alpha_i$ were to be fitted to $\gamma p \to K^+ \pi \Sigma$ 
experimental data. In this way one allows to fine tune the theoretical chiral 
unitary inspired model, incorporating in an effective way possible contributions 
of higher order terms,  and extract from experiment an accurate position for the 
two $\Lambda(1405)$ poles and the actual shape of the meson-baryon scattering 
amplitudes.

On the other hand, the $G_i$ function in Eq.~({\ref{eq:BS}) (as defined in Eq.~(\ref{eq:bonnG}))
can be regularized either with a three--momentum cutoff or with dimensional 
regularization in 
terms of subtraction constants, $a_i$, one for each meson-baryon channel. In 
Refs.~\cite{Roca:2013av,Roca:2013cca} these parameters were also allowed to vary 
slightly substituting them by  $a_{KN}\to \beta_1 a_{KN}$, $a_{\pi\Sigma}\to \beta_2 
a_{\pi\Sigma}$  with  $a_{KN} = -1.84$, $a_{\pi\Sigma}=-2$ \cite{Oset:2001cn,cola}. 
All in all, there are only five $\alpha_i$, $\beta_i$, parameters needed in the 
present work. Their values are taken from 
Table~I in \cite{Roca:2013cca}.

%$\alpha_{11}=1.037$, $\alpha_{12}=1.466$,  $ \alpha_{22}=1.668$, $\beta_1=1.187$ and $\beta_2=0.722$.
%
%\mm{\sout{With this solution we obtain  the scattering amplitudes shown in Fig.~\ref{fi%g:t_MMMM0fit}. }
%MM: I think figure is too much. I also would prefer to refer to matrix $C$ in eq. 9 and% 10 just saying 
%this is the WT term modified slightly. All details are given in the original publicatio%n, but this is 
%decision of the authors of the model :-).} 
When looking for poles in the second Riemann sheet of the complex energy plane, 
the amplitudes of this model provide the $\Lambda(1405)$ pole positions at $1352-48i$~MeV, 
and $1419-29i$~MeV. Note that the parameters do not differ much from one, as 
would be expected if reality is not far from the predictions of the
chiral unitary theory. In the $\bar K N\to\bar K N$ amplitude the highest pole 
is more pronounced. The 
$\pi \Sigma\to \pi \Sigma$ amplitude picks more the lowest pole while in 
the  $\bar K N\to\pi \Sigma$
amplitude a more balanced mixture between both poles is visible but with a larger weight of the highest
one. (See, for instance, Fig.~6 in ref.~\cite{Roca:2013cca}). All this is reminiscent of the fact that the highest pole couples 
dominantly to $\bar K N$ and the
lowest pole to $\pi \Sigma$, see Table~II in Ref.~\cite{Roca:2013cca}.

%\mm{\sout{It is worth noting that the poles are not included as explicit degrees of fre%edom, but they appear 
%dynamically from the highly non-linear dynamics of the meson-baryon unitarization proce%dure. Furthermore, 
%the unitarization procedure provides not only the pole positions but the actual line-sh%apes of the amplitudes 
%in the real axis, which are very far from being Breit-Wigner, and of crucial importance% in actual experimental 
%observables like the invariant mass distributions of concern in the present work.}}

%%%%%%%%%%%%%%%%%%%%%%%%%%%%%%%%%%%%%%%%%%%%%%%%%%%%%%%%%%%%%%%%%%
\section{Results}\label{sec:results}
%%%%%%%%%%%%%%%%%%%%%%%%%%%%%%%%%%%%%%%%%%%%%%%%%%%%%%%%%%%%%%%%%%

%%%%%%%%%%%%%%%%%%%%%%%%%%%%%%%%%%%%%%%%%%%%%%%%%%%%
\begin{figure}[t]
\begin{center}
\includegraphics[width=\linewidth]{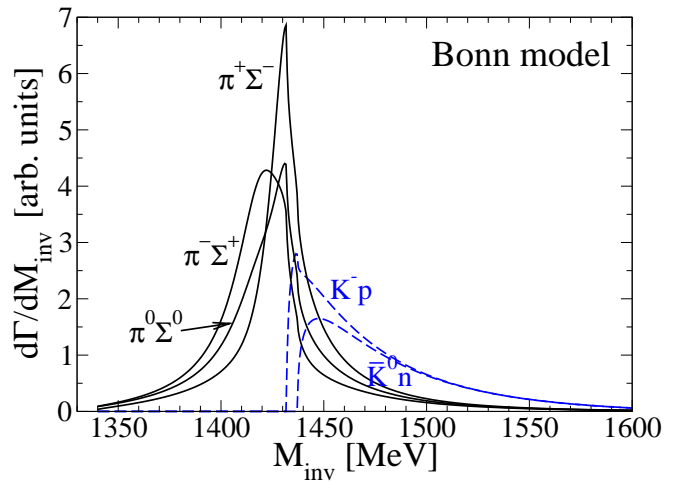}
\caption{Invariant mass distribution within the Bonn model 
considering isospin breaking. }
\label{fig:fig2PHYSMM}
\end{center}
\end{figure}
%%%%%%%%%%%%%%%%%%%%%%%%%%%%%%%%%%%%%%%%%%%%%%%%%%%%%

After having set up the framework, we present here our predictions for the  
$\pi\Sigma$ and $\bar K N$ invariant mass distributions from the $\Lambda_b$ decay. 
As mentioned before, one of the important features\footnote{Which is unfortunately  not very common in such studies.}
of the present study is quantification of  the theoretical uncertainties, due to 
different meson-baryon models. To make  this comparison more meaningful, the 
trivial sources of differences must be  studied first, such as  isospin symmetry. 
The latter is implemented
in  the MV-model by construction, while it is broken explicitly in the Bonn model.  The
isospin-breaking in the Bonn model arises naturally due to chiral potential  of the
next-to-leading order.   All particle masses are considered to be the physical ones, see the
discussion  in Refs.~\cite{Bruns:2010sv,Mai:2012dt,Mai:2014xna}. In
Fig.~\ref{fig:fig2PHYSMM}  we show the results for the Bonn model considering 
the explicit isospin-breaking. This provides the order of the
correction  for the subsequent figures if one considers  isospin-breaking. 
In the following we will consider the isospin-symmetric case for simplicity and  
to ease the comparison with the MV-model.

%%%%%%%%%%%%%%%%%%%%%%%%%%%%%%%%%%%%%%%%%%%%%%%%%%%%%
\begin{figure*}[t]
\includegraphics[width=0.8\linewidth]{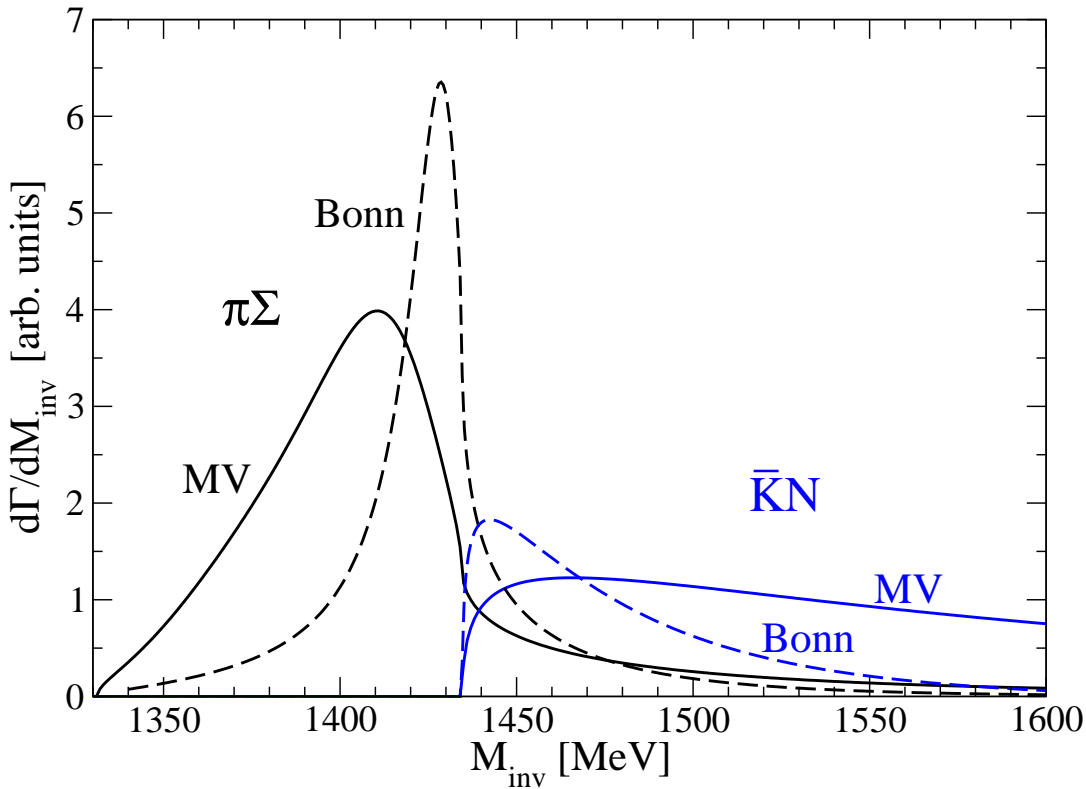}
\caption{Results for the $\pi\Sigma$ and $\bar KN$ invariant mass distributions 
for the $\Lambda_b \to J/\psi\,\pi\Sigma$ and $\Lambda_b \to J/\psi\,\bar KN$ decays, 
respectively, for both models considered in the present work.}\label{fig:fig1}
\end{figure*}
%%%%%%%%%%%%%%%%%%%%%%%%%%%%%%%%%%%%%%%%%%%%%%%%%%%%%

In Fig.~\ref{fig:fig1} we show the final results for both the Bonn and MV models.
In the $\pi\Sigma$ final state channel the peak of the $\Lambda(1405)$ is  clearly 
visible.
In fact, this is mostly due to the higher mass pole of  the $\Lambda(1405)$ since the
contribution proportional to $t_{\bar K N,\pi\Sigma}$  of Eq.~(\ref{eqn:fullamplitude}) is
the dominant one.  The difference in the $\pi\Sigma$ mass distribution between both 
models is
reminiscent of the fact that, as explained above, the Bonn model gets a narrower 
($24\mev$)
highest $\Lambda(1405)$ pole than the MV model ($58\mev$).

In the $\bar K N$ final state,  the dominant contribution comes from the
part proportional to $t_{\bar K N,\bar K N}$  which again is more sensitive to the higher
mass $\Lambda(1405)$ pole. However,  in this channel only the effect of the tail of the
resonance is visible since  the threshold of the $\bar K N$ mass distribution is located
above the position  of the $\Lambda(1405)$ peak. Nevertheless, that tail is enough to
provide  a high strength close to the threshold, what makes the line shape of the  $\bar K N$
invariant mass distribution  to be very different from just  a phase-space 
distribution. The
dependence on the choice of the model in this channel is due to the fact that
 the highest pole is
slightly closer to threshold in the  Bonn model compared to the MV one. Because of this
feature, the Bonn model produces  a narrower bump close to $\bar K N$ invariant mass
threshold than the MV one.  This observable is then very sensitive to the exact 
position of
the resonance  pole, due to the proximity between the threshold and the pole. As 
mentioned in  the introduction, different reactions can reflect different weights 
for both  poles of
the $\Lambda(1405)$ resonance, depending on the particular production  dynamics. In the
present case, the highest pole is the one that shows up dominantly.

On the other hand, the agreement in Fig.~\ref{fig:fig1} of the results 
between the MV and Bonn models is remarkable, given their theoretical 
differences and fitting strategies as explained before. Nonetheless we can 
regard the difference between the models as the main source of the 
theoretical uncertainty. 

While the overall normalization of the invariant mass distributions is unknown,
the shape and the 
ratio between the  $\pi\Sigma$ and $\bar K N$ distributions is 
unchanged and it is a genuine prediction of the present work.
 Indeed, the ratio between the maximum values of 
the  $\pi\Sigma$ and $\bar K N$ distribution  is 3.3 for the MV and 3.5 for 
the Bonn model. 
The value of that ratio as well as the shape of the distributions are then genuine 
predictions of the chiral unitary approach. As already stated, the differences 
between the 
different curves can be considered  as an estimation of the theoretical uncertainty. 
In conclusion, Fig.~\ref{fig:fig1} serves to predict the  invariant mass 
distributions of either $\pi\Sigma$ or $\bar K N$,  once the absolute 
normalization of the mass distribution of the other channel has been measured. 
For instance, if the LHCb \cite{Aaij:2014zoa} and CDF \cite{Aaltonen:2014vra} 
collaboration were  to measure the $K^-p$ mass distribution in the 
$\Lambda_b \to J/\psi ~ K^- p$ decay, then the shape should agree with the 
prediction of Fig.~\ref{fig:fig1} and once normalized, 
the $\bar K N$ and $\pi\Sigma$ distributions would be given both in size and shape.

%%%%%%%%%%%%%%%%%%%%%%%%%%%%%%%%%%%%%%%%%%%%%%%%%%%%%%%%%%%%%%%%%%
\section{Summary}
%%%%%%%%%%%%%%%%%%%%%%%%%%%%%%%%%%%%%%%%%%%%%%%%%%%%%%%%%%%%%%%%%%

We have carried out a theoretical study of the $\Lambda(1405)$ production  
in the $\Lambda_b \to J/\psi\,\pi\Sigma$ and $\Lambda_b \to J/\psi \,\bar K N$ decays.  The initial weak production at the level of quarks to give a $c\bar c$ for  
the $J/\Psi$ and three quarks
$uds$ is the same for both channels and then irrelevant in the relative ratio. The
hadronization of the $uds$ into the different meson-baryon channels is then 
implemented and
the different channels are related using suitable $SU(3)$ arguments.

The key point of the chiral unitary models  is that the
$\Lambda(1405)$ comes out as dynamically generated. Actually, two poles are 
predicted for
this resonance. Accordingly, we implement the final state
interaction of the meson-baryon pair, using two different theoretical models
\cite{Roca:2013cca,Mai:2014xna}. The MV-model \cite{Roca:2013av,Roca:2013cca} 
uses as the
kernel of the unitarization procedure the lowest order meson-baryon chiral Lagrangian
slightly modified to fit photoproduction data. On the other hand, the Bonn model
\cite{Mai:2014xna} includes in the kernel from higher order 
meson-baryon Lagrangians fitted to photoproduction and meson-baryon cross section data.

The $\Lambda(1405)$ resonant shape is clearly  visible in the $\pi\Sigma$ mass 
distribution and its tail distorts considerably the $\bar K N$ distribution in 
spite of the pole being below
the $\bar K N$ threshold. This particular decay is mostly influenced by the 
higher mass pole
of  the $\Lambda(1405)$ resonance. Therefore this decay is specially suited 
to study the
properties of the high mass $\Lambda(1405)$ resonance both theoretically and
experimentally. 

The results for both theoretical models  used in the present work are 
remarkably similar and their differences can be considered as the theoretical 
uncertainty of this calculation. The line
shapes of the $\pi\Sigma$ and $\bar K N$ distributions and their relative strengths are
predictions of this model which could be compared to future experimental measurements
amenable to study the $\Lambda(1405)$ resonance in this decay.

%%%%%%%%%%%%%%%%%%%%%%%%%%%%%%%%%%%%%%%%%%%%%%%%%%%%%%%%%%%%%%%%%%
\section*{Acknowledgments}
This work is partly supported by the Spanish Ministerio de Economia y Competitividad and
European FEDER funds under the contract number FIS2011-28853-C02-01 and FIS2011-28853-C02-02,
and the Generalitat Valenciana in the program Prometeo II-2014/068. We acknowledge the
support of the European Community-Research Infrastructure Integrating Activity Study of
Strongly Interacting Matter (acronym HadronPhysics3, Grant Agreement n. 283286) under the
Seventh Framework Programme of EU. Work supported in part  by the DFG and NSFC through
funds provided to the Collaborative Research Center CRC~110 ``Symmteries and the Emergence of
Structure in QCD'', and by the DFG (TR~16).
E.~O. wishes to thank Tetsuo Hyodo for useful discussions

%%%%%%%%%%%%%%%%%%%%%%%%%%%%%%%%%%%%%%%%%%%%%%%%%%%%%%%%%%%%%%%%%%


\begin{thebibliography}{999}
\bibitem{Chau:1982da}
  L.~L.~Chau,
    Phys.\ Rept.\  {\bf 95}, 1 (1983).
    

\bibitem{Chau:1987tk}
  L.~L.~Chau and H.~Y.~Cheng,
    Phys.\ Rev.\ D {\bf 36}, 137 (1987).
    

\bibitem{Cheng:2010vk} 
  H.~Y.~Cheng and C.~W.~Chiang,
    Phys.\ Rev.\ D {\bf 81}, 074031 (2010).
    

\bibitem{Aaij:2011fx}
  R.~Aaij {\it et al.}  [LHCb Collaboration],
    Phys.\ Lett.\ B {\bf 698}, 115 (2011).
 

\bibitem{Li:2011pg}
  J.~Li {\it et al.}  [Belle Collaboration],
    Phys.\ Rev.\ Lett.\  {\bf 106}, 121802 (2011).
  

\bibitem{Aaij:2013zpt}
  R. Aaij {\it et al.}  [LHCb Collaboration],
    Phys.\ Rev.\ D {\bf 87}, 052001 (2013).
    

\bibitem{stone}
  S.~Stone and L.~Zhang,
    Phys.\ Rev.\ Lett.\  {\bf 111}, 062001 (2013).
    

\bibitem{weihong} 
  W.~H.~Liang and E.~Oset,
    Phys.\ Lett.\ B {\bf 737}, 70 (2014).
    

\bibitem{npa}
  J.~A.~Oller and E.~Oset,
    Nucl.\ Phys.\ A {\bf 620}, 438 (1997)
  [Erratum-ibid.\ A {\bf 652}, 407 (1999)].
    

\bibitem{ramonet}
  J.~A.~Oller, E.~Oset and J.~R.~Pelaez,
    Phys.\ Rev.\ D {\bf 59}, 074001 (1999)
  [Erratum-ibid.\ D {\bf 60}, 099906 (1999)]
  [Erratum-ibid.\ D {\bf 75}, 099903 (2007)].
    
  

\bibitem{kaiser}
N.~Kaiser,
Eur.\ Phys.\ J.\ A {\bf 3}, 307 (1998).

\bibitem{markushin}
M.~P.~Locher, V.~E.~Markushin and H.~Q.~Zheng,
Eur.\ Phys.\ J.\ C {\bf 4}, 317 (1998).

\bibitem{juanito}
J.~Nieves and E.~Ruiz Arriola,
Nucl.\ Phys.\ A {\bf 679}, 57 (2000).

\bibitem{rios}
J.~R.~Pelaez and G.~Rios,
Phys.\ Rev.\ Lett.\ {\bf 97}, 242002 (2006). 

\bibitem{Meissner:2000bc}
  U.-G.~Mei{\ss}ner and J.~A.~Oller,
    Nucl.\ Phys.\ A {\bf 679} (2001) 671.

\bibitem{Gardner:2001gc} 
  S.~Gardner and U.-G.~Mei{\ss}ner,
    Phys.\ Rev.\ D {\bf 65}, 094004 (2002).

\bibitem{Doring:2013wka} 
  M.~D\"oring, U.-G.~Mei{\ss}ner and W.~Wang,
    JHEP {\bf 1310}, 011 (2013).

\bibitem{Meissner:2013pba} 
  U.-G.~Mei{\ss}ner and W.~Wang,
    JHEP {\bf 1401}, 107 (2014).

\bibitem{Bayar:2014qha} 
  M.~Bayar, W.~H.~Liang and E.~Oset,
    Phys.\ Rev.\ D {\bf 90}, no. 11, 114004 (2014).
    

\bibitem{daid} 
  J.~J.~Xie, L.~R.~Dai and E.~Oset,
    arXiv:1409.0401 [hep-ph]. Phys. Lett. B, in print.
    

\bibitem{xievec} 
  J.~J.~Xie and E.~Oset,
    Phys.\ Rev.\ D {\bf 90}, no. 9, 094006 (2014).
    

\bibitem{xiebd} 
  W.~H.~Liang, J.~J.~Xie and E.~Oset,
    arXiv:1501.00088 [hep-ph].
  

\bibitem{marina} 
  M.~Albaladejo, M.~Nielsen and E.~Oset,
    arXiv:1501.03455 [hep-ph].
  

\bibitem{fernando} 
  F.~S.~Navarra, M.~Nielsen, E.~Oset and T.~Sekihara,
    arXiv:1501.03422 [hep-ph].
  

\bibitem{Abazov:2007sf} 
  V.~M.~Abazov {\it et al.}  [D0 Collaboration],
    Phys.\ Rev.\ Lett.\  {\bf 99}, 142001 (2007).
    

\bibitem{Aad:2014iba} 
  G.~Aad {\it et al.}  [ATLAS Collaboration],
    Phys.\ Rev.\ D {\bf 89}, no. 9, 092009 (2014).
    

\bibitem{Aaij:2014zoa} 
  R.~Aaij {\it et al.}  [LHCb Collaboration],
    JHEP {\bf 1407}, 103 (2014).
    

\bibitem{Aaltonen:2014vra} 
  T.~A.~Aaltonen {\it et al.}  [CDF Collaboration],
    Phys.\ Rev.\ Lett.\  {\bf 113}, no. 24, 242001 (2014).
    

\bibitem{Dalitz:1960du} 
  R.~H.~Dalitz and S.~F.~Tuan,
    Annals Phys.\  {\bf 10}, 307 (1960).
    

\bibitem{Dalitz:1967fp} 
  R.~H.~Dalitz, T.~C.~Wong and G.~Rajasekaran,
    Phys.\ Rev.\  {\bf 153}, 1617 (1967).
    

\bibitem{Veit:1984an} 
  E.~A.~Veit, B.~K.~Jennings, R.~C.~Barrett and A.~W.~Thomas,
    Phys.\ Lett.\ B {\bf 137}, 415 (1984).
    
 

\bibitem{Kaiser:1995eg}
  N.~Kaiser, P.~B.~Siegel and W.~Weise,
    Nucl.\ Phys.\  A {\bf 594}, 325 (1995).
    

\bibitem{Kaiser:1996js} 
  N.~Kaiser, T.~Waas and W.~Weise,
    Nucl.\ Phys.\ A {\bf 612}, 297 (1997).
  

\bibitem{Oset:1998it}
  E.~Oset and A.~Ramos,
    Nucl.\ Phys.\  A {\bf 635}, 99 (1998).
  

\bibitem{Oller:2000fj}
  J.~A.~Oller and U.-G.~Mei{\ss}ner,
       Phys.\ Lett.\  B {\bf 500}, 263 (2001).
    
   

\bibitem{Lutz:2001yb}
  M.~F.~M.~Lutz and E.~E.~Kolomeitsev,
       Nucl.\ Phys.\  A {\bf 700}, 193 (2002).
    
   

\bibitem{Oset:2001cn}
  E.~Oset, A.~Ramos and C.~Bennhold,
    Phys.\ Lett.\  B {\bf 527}, 99 (2002)
  [Erratum-ibid.\  B {\bf 530}, 260 (2002)].
    
  

\bibitem{Hyodo:2002pk}
  T.~Hyodo, S.~I.~Nam, D.~Jido and A.~Hosaka,
       Phys.\ Rev.\  C {\bf 68}, 018201 (2003).
    
    

\bibitem{cola}
  D.~Jido, J.~A.~Oller, E.~Oset, A.~Ramos, U.-G.~Mei{\ss}ner,
    Nucl.\ Phys.\  {\bf A725}, 181-200 (2003).
  

\bibitem{GarciaRecio:2002td} 
  C.~Garcia-Recio, J.~Nieves, E.~Ruiz Arriola and M.~J.~Vicente Vacas,
    Phys.\ Rev.\ D {\bf 67}, 076009 (2003).
  
  

\bibitem{GarciaRecio:2005hy} 
  C.~Garcia-Recio, J.~Nieves and L.~L.~Salcedo,
    Phys.\ Rev.\ D {\bf 74}, 034025 (2006).
  

\bibitem{Borasoy:2005ie}
  B.~Borasoy, R.~Nissler and W.~Weise,
    Eur.\ Phys.\ J.\  A {\bf 25}, 79 (2005).
    
  

\bibitem{Oller:2006jw}
  J.~A.~Oller,
    Eur.\ Phys.\ J.\  A {\bf 28}, 63 (2006).
  

\bibitem{Borasoy:2006sr}
  B.~Borasoy, U.-G.~Mei{\ss}ner and R.~Nissler,
    Phys.\ Rev.\  C {\bf 74}, 055201 (2006).
    

\bibitem{hyodonew} 
  Y.~Ikeda, T.~Hyodo and W.~Weise,
    Nucl.\ Phys.\ A {\bf 881}, 98 (2012).
    

\bibitem{Mai:2012dt} 
  M.~Mai and U.-G.~Mei{\ss}ner,
    Nucl.\ Phys.\ A {\bf 900}, 51  (2013)
  [arXiv:1202.2030 [nucl-th]].
  

\bibitem{Fink:1989uk} 
  P.~J.~Fink, Jr., G.~He, R.~H.~Landau and J.~W.~Schnick,
    Phys.\ Rev.\ C {\bf 41}, 2720 (1990).
    
 

\bibitem{Thomas:1973uh} 
  D.~W.~Thomas, A.~Engler, H.~E.~Fisk and R.~W.~Kraemer,
    Nucl.\ Phys.\ B {\bf 56}, 15 (1973).
   
\bibitem{Hemingway:1984pz} 
  R.~J.~Hemingway,
    Nucl.\ Phys.\ B {\bf 253}, 742 (1985).
  
\bibitem{Niiyama:2008rt} 
  M.~Niiyama, H.~Fujimura, D.~S.~Ahn, J.~K.~Ahn, S.~Ajimura, H.~C.~Bhang, T.~H.~Chang and W.~C.~Chang {\it et al.},
    Phys.\ Rev.\ C {\bf 78}, 035202 (2008).
  
\bibitem{prakhov} 
  S.~Prakhov {\it et al.}  [Crystall Ball Collaboration],
    Phys.\ Rev.\ C {\bf 70}, 034605 (2004).
    
\bibitem{Moriya:2012zz} 
  K.~Moriya {\it et al.}  [CLAS Collaboration],
    AIP Conf.\ Proc.\  {\bf 1441}, 296 (2012).
   
\bibitem{Moriya:2013eb} 
  K.~Moriya {\it et al.}  [CLAS Collaboration],
    Phys.\ Rev.\ C {\bf 87}, no. 3, 035206 (2013).
      
\bibitem{Zychor:2007gf} 
  I.~Zychor, M.~B\"uscher, M.~Hartmann, A.~Kacharava, I.~Keshelashvili, A.~Khoukaz, V.~Kleber and V.~Koptev {\it et al.},
    Phys.\ Lett.\ B {\bf 660}, 167 (2008).
    
\bibitem{fabbietti} 
  G.~Agakishiev {\it et al.}  [HADES Collaboration],
    Phys.\ Rev.\ C {\bf 87}, no. 2, 025201 (2013).
    
\bibitem{magaslam} 
  V.~K.~Magas, E.~Oset and A.~Ramos,
    Phys.\ Rev.\ Lett.\  {\bf 95}, 052301 (2005).
    
\bibitem{Braun:1977wd} 
  O.~Braun, H.~J.~Grimm, V.~Hepp, H.~Strobele, C.~Thol, T.~J.~Thouw, D.~Capps and F.~Gandini {\it et al.},
    Nucl.\ Phys.\ B {\bf 129}, 1 (1977).
   
\bibitem{sekihara} 
  D.~Jido, E.~Oset and T.~Sekihara,
    Eur.\ Phys.\ J.\ A {\bf 42}, 257 (2009).
  
\bibitem{Miyagawa:2012xz}
  K.~Miyagawa and J.~Haidenbauer,
    Phys.\ Rev.\ C {\bf 85} (2012) 065201.

\bibitem{Jido:2012cy}
  D.~Jido, E.~Oset and T.~Sekihara,
    Eur.\ Phys.\ J.\ A {\bf 49} (2013) 95.
  
\bibitem{Ren:2015bsa} 
  X.~Ren, E.~Oset, L.~Alvarez-Ruso and M.~J.~V.~Vacas,
    arXiv:1501.04073 [hep-ph].
  
\bibitem{Roca:2013av} 
  L.~Roca and E.~Oset,
    Phys.\ Rev.\ C {\bf 87}, no. 5, 055201 (2013).
    
\bibitem{Roca:2013cca} 
  L.~Roca and E.~Oset,
    Phys.\ Rev.\ C {\bf 88}, no. 5, 055206 (2013).
    
\bibitem{Mai:2014xna} 
  M.~Mai and U.-G.~Mei{\ss}ner,
    arXiv:1411.7884 [hep-ph], accepted for publication in Eur. Phys. J. {\bf A} (2015).
    
\bibitem{Hyodo:2011js} 
  T.~Hyodo and M.~Oka,
    Phys.\ Rev.\ C {\bf 84}, 035201 (2011).
    
\bibitem{Gutsche:2013oea} 
  T.~Gutsche, M.~A.~Ivanov, J.~G.~K\"orner, V.~E.~Lyubovitskij and P.~Santorelli,
    Phys.\ Rev.\ D {\bf 88}, no. 11, 114018 (2013).
    
\bibitem{Albertus:2014gba} 
  C.~Albertus, E.~Hernandez, C.~Hidalgo-Duque and J.~Nieves,
    Phys.\ Lett.\ B {\bf 738}, 144 (2014).
    
\bibitem{Jia:2012vt} 
  Y.~Jia, F.~Jugeau and L.~Oliver,
    Phys.\ Rev.\ D {\bf 86}, 014002 (2012).
    
\bibitem{Bramon:1992kr}
  A.~Bramon, A.~Grau and G.~Pancheri,
    Phys.\ Lett.\ B {\bf 283}, 416 (1992).
   
\bibitem{close_book}
F. E Close
 {\it An introduction to quarks and partons}.
Academic Press, 1979 

\bibitem{Bruns:2010sv}
  P.~C.~Bruns, M.~Mai and U.-G.~Mei{\ss}ner,
    Phys.\ Lett.\ B {\bf 697} (2011) 254.
      
\bibitem{Mai:2012wy}
  M.~Mai, P.~C.~Bruns and U.-G.~Mei{\ss}ner,
    Phys.\ Rev.\ D {\bf 86} (2012) 094033.
    
\bibitem{Frink:2004ic}
  M.~Frink and U.-G.~Mei{\ss}ner,
      JHEP {\bf 0407} (2004) 028.

\bibitem{Hyodo:2006kg}
T.~Hyodo, D.~Jido, and A.~Hosaka,
\newblock Phys. Rev. {\bf D75}, 034002 (2007).

\bibitem{dobado-pelaez}
A.~Dobado and J.~R.~Pel\'aez,
Phys.\ Rev.\ D {\bf 56} (1997) 3057.

\bibitem{Oller:1998zr}
J.~A.~Oller and E.~Oset,
Phys.\ Rev.\ D {\bf 60} (1999) 074023.

\bibitem{Hyodo:2003qa}
T.~Hyodo, S.~I. Nam, D.~Jido, and A.~Hosaka,
\newblock Prog. Theor. Phys. {\bf 112}, 73 (2004).

\bibitem{Bernard:1995dp}
  V.~Bernard, N.~Kaiser and U.-G.~Mei{\ss}ner,
    Int.\ J.\ Mod.\ Phys.\  E {\bf 4} (1995) 193.
  

\end{thebibliography}
\end{document}